\documentclass[prb,twocolumn,superscriptaddress]{revtex4-1}
\usepackage{amsfonts,amsmath,graphicx,amssymb}
\pdfoutput=1

\newcommand{\om}{\omega}
\newcommand{\Qv}{\mathbf{Q}} 

\begin{document}
\title{Resilience of $d$-wave superconductivity to nearest-neighbor repulsion}

\author{D. S\'en\'echal}
\affiliation{D\'epartement de physique and Regroupement qu\'eb\'ecois sur les mat\'eriaux de pointe, Universit\'e de Sherbrooke, Sherbrooke, Qu\'ebec, Canada J1K 2R1}
\author{A. Day}
\affiliation{D\'epartement de physique and Regroupement qu\'eb\'ecois sur les mat\'eriaux de pointe, Universit\'e de Sherbrooke, Sherbrooke, Qu\'ebec, Canada J1K 2R1}
\author{V. Bouliane}
\affiliation{D\'epartement de physique and Regroupement qu\'eb\'ecois sur les mat\'eriaux de pointe, Universit\'e de Sherbrooke, Sherbrooke, Qu\'ebec, Canada J1K 2R1}
\author{A.-M. S. Tremblay}
\affiliation{D\'epartement de physique and Regroupement qu\'eb\'ecois sur les mat\'eriaux de pointe, Universit\'e de Sherbrooke, Sherbrooke, Qu\'ebec, Canada J1K 2R1}
\affiliation{Canadian Institute for Advanced Research, Toronto, Ontario, Canada , M5G 1Z8}

\begin{abstract}
Many theoretical approaches find $d$-wave superconductivity in the prototypical one-band Hubbard model for high-temperature superconductors. At strong-coupling ($U\geq W$, where $U$ is the on-site repulsion and $W=8t$ the bandwidth) pairing is controlled by the exchange energy $J=4t^2/U$. One may then surmise, ignoring retardation effects, that near-neighbor Coulomb repulsion $V$ will destroy superconductivity when it becomes larger than $J$, a condition that is easily satisfied in cuprates for example. 
Using Cellular Dynamical Mean-Field theory with an exact diagonalization solver for the extended Hubbard model, we show that pairing {\it at strong coupling} is preserved, even when $V\gg J$, as long as $V\lesssim U/2$. While at weak coupling $V$ always reduces the spin fluctuations and hence $d$-wave pairing, at strong coupling, in the underdoped regime, the increase of $J=4t^2/(U-V)$ caused by $V$ increases binding at low frequency while the pair-breaking effect of $V$ is pushed to high frequency. These two effects compensate in the underdoped regime, in the presence of a pseudogap. 
While the pseudogap competes with superconductivity, the proximity to the Mott transition that leads to the pseudogap, and retardation effects, protect $d$-wave superconductivity from $V$. 
\end{abstract}

\maketitle


\section{Introduction}

The existence of $d$-wave superconductivity in the one-band two-dimensional Hubbard model has been established through a variety of theoretical methods at both weak~\cite{kohn:1965,Fay:1968,Baranov:1992,Hlubina:1999,Mraz:2003, Raghu:2010,Beal-Monod:1986,Bickers_dwave:1989,ScalapinoThread:2010,ZanchiSchulz:1998,ZanchiSchulz:2000,Halboth:2000a,Halboth:2000b,Honerkamp:2001a,Honerkamp:2001,HonerkampSalmhoferTflow:2001,HonerkampSalmhoferRice:2002,Tsai:2001,MetznerReview:2012,Kyung:2003,Bickers_dwave:1989,MonthouxScalapino:1994,DahmTewordt:1995,ManskeEremin:2003,AbanovChubukovNorman:2008,Hassan:2008} and strong coupling~\cite{Giamarchi:1991,Paramekanti:2001,Paramekanti:2004,AndersonVanilla:2004,PathakShenoyRanderiaTrivedi:2009,Kotliar:1988,LeeRMP:2006}, in other words for one-site interaction $U$ either much smaller or much larger than the bandwidth $W$. Generalizations of Dynamical Mean-Field Theory are particularly suited for the strong coupling limit, but they are also an excellent guide to the physics at weak to intermediate coupling.~\cite{Maier:2000a,maier_d:2005,Senechal:2005,AichhornAFSC:2006,Aichhorn:2007,Haule:2007,Kancharla:2008,Kyung:2009,GullParcollet:2012,SordiSuperconductivityPseudogap:2012}, These calculations suggest that pairing is maximized at intermediate coupling, where the on-site interaction $U$ is of order the bandwidth $W=8t$. Some non-perturbative calculations based on weak coupling ideas even agree at intermediate coupling~\cite{LTP:2006} with strong-coupling based approaches.   

In all these approaches, spin fluctuations with either an antiferromagnetic or a singlet character~\cite{Haule:2007,Anderson:1987,Anderson:2007} have been argued to drive the pairing. These spin fluctuations result from the presence of an on-site Coulomb repulsion $U$. At strong coupling, the characteristic energy scale of these fluctuations, the exchange interaction $J$, is given by $4t^2/U$, and the $d$-wave gap symmetry adopted by the Cooper pairs allows them to avoid the direct effect of the on-site repulsion $U$. 

Little attention has been paid so far to the effect of the nearest-neighbor Coulomb repulsion (or extended Hubbard interaction) $V$ that we expect to be detrimental to $d$-wave superconductivity. Roughly speaking, in a simple BCS picture that does not take retardation into account, we expect the effective interaction to be the difference $J-V$. In ordinary phonon-mediated superconductivity, the repulsion $V$ is replaced by a smaller pseudopotential $V_c$ to account for the fact that binding occurs at low frequencies through phonons while the Coulomb interaction acts over a broad energy scale. This so-called Anderson-Morel mechanism~\cite{AndersonMorel:1962,Scalapino:1966}, leads to the following estimate for the Coulomb pseudopotential $V_c=V/(1+N(0)V\ln(E_F/\om_D))$ where the Debye frequency is $\om_D$ and the Fermi energy $E_F$. One expects that in strongly-correlated superconductivity, the ratio $E_F/\om_D$ must be replaced by a number closer to unity in which case this mechanism would no-longer be effective and superconductivity should disappear as soon as  $V > J$. This issue is crucial to understand high-temperature superconductors since that condition, or the weaker condition $V > \Delta_s$ with $\Delta_s$ the spin gap, is likely to be satisfied in these materials. From the value of the near-neighbor Coulomb interaction with a relative dielectric constant of order 10 we estimate $V\approx 400$ meV while $J\approx 130$ meV~\cite{HaydenJHeisenberg:1991}. 

So far, it has been shown using a variational wave-function approach for the $t-U-J-V$ model at strong coupling that superconductivity persists as long as~\cite{PlekhanovSorella:2003} $3J/4 > \delta^2(V-J/4)$ where $\delta$ is the doping.  A large-N calculation gives superconductivity at least up to $V=2J$,~\cite{Zhou-tJV:2004} while Density Matrix Renormalization Group (DMRG) calculations on Hubbard or $t-J$ ladders~\cite{NoackWhiteSupra:1995,NoackBulutScalapino:1997,ArrigoniHarjuKivelson:2002} suggest that pairing can survive up to $V\approx 4J$.~\cite{Raghu:2012} At very small coupling ($U\ll W$), it has been argued~\cite{Raghu:2012} that pairing is destroyed as soon as $V\geq U(U/W)$. This weak coupling bound is close to the result of a FLEX calculation.~\cite{OnariExtended:2004}

Here we show that $d$-wave superconductivity in the one-band two-dimensional Hubbard model at strong coupling is in fact more robust than expected. Even for $V\gg J$, as long as the inequality $V< U/2$ is satisfied, superconductivity persists. This illustrates differences between pairing at weak and at strong coupling, especially in the pseudogap regime. The resilience of $d$-wave superconductivity to $V$ can be traced to the increase in the effective $J$ caused by $V$ at strong coupling, i.e. when at half-filling the system is a Mott insulator. This increase in $J$ is visible in the dynamics of both the spin susceptibility and the Cooper pair Green's function which is enhanced by $V$ at low frequency. The pair-breaking effect of $V$ manifests itself at higher frequency. This leads overall to a sizeable range of values of $V$ where the order parameter is essentially independent of $V$ in the underdoped regime where a pseudogap appears. Recall that at strong coupling, the pseudogap extends up to optimal doping, a sign that Mott physics extends well beyond half-filling~\cite{Sordi:2010}.

We use Cellular-Dynamical Mean-Field Theory that allows us to study the strong-coupling limit and allows one to take into account both $J$ and $V$ as well as the effect of retardation.    



We first present the method and model and conclude with a discussion after the presentation of the results obtained from large scale numerical calculations.

\section{Model and method}

We start from the one-band extended Hubbard model on a square lattice, 
\begin{equation}\label{eq:hamil}
H =-\sum_{i,j,\sigma}t_{ij}c_{i,\sigma}^{\dagger}c_{j,\sigma}
+U\sum_{i}n_{i\uparrow}n_{i\downarrow}  + V \sum_{\langle i,j\rangle}n_in_j
\end{equation}
where $c_{i,\sigma}^{(\dagger)}$ is the destruction (creation) operator for an electron of spin $\sigma$ at site $i$ and  $n_{i\sigma}=c_{i,\sigma}^{\dagger}c_{i,\sigma}$ is the corresponding number operator ($n_i=n_{i\uparrow}+n_{i\downarrow}$). 
We assume a band structure close to that of YBa$_2$Cu$_3$O$_7$, with nearest-neighbor hopping $t$ set to unity, diagonal hopping $t'=-0.3$ and third-neighbor hopping $t''=0.2$, unless otherwise indicated.
This model is solved with a 4 site plaquette Cellular Dynamical Mean Field Theory (CDMFT) at $T=0$ using an exact diagonalization solver~\cite{Caffarel:1994,Kancharla:2008,Senechal:2011,LiebschIshidaBathReview:2012} that allows us to obtain real-time quantities without analytic continuation. This approach has been used to reveal the presence of $d$-wave superconductivity in the one-band Hubbard model,\cite{Kancharla:2008} and to study the pairing dynamics and retardation when $V=0$.~\cite{Kyung:2009} The results obtained for $d$-wave superconductivity with the plaquette are essentially identical to those obtained with larger clusters.~\cite{GullParcollet:2012}

\begin{figure}
\centering{
\includegraphics[scale=1.1]{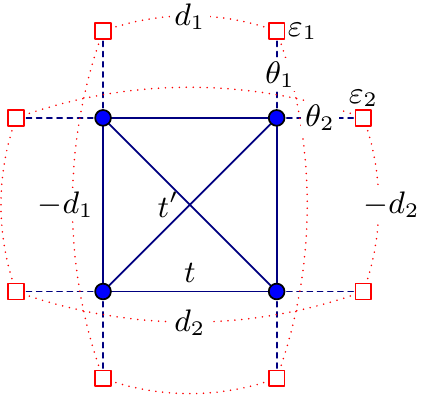}}
\caption{Cluster and bath parametrization used in this work. Hopping terms are shown by full lines, bath hybridization by dashed lines and bath pairing terms by dotted lines.}
\label{fig:cluster}
\end{figure}

As in previous studies of the 2D Hubbard model with this approach~\cite{Kancharla:2008}, the plaquette is hybridized with a set of 8 bath orbitals, as illustrated in Fig.~\ref{fig:cluster}, with 6 parameters to be determined through the CDMFT self-consistency relation: the hybridizations $\theta_{1,2}$ between the bath and the cluster, the bath energies $\varepsilon_{1,2}$ and the bath $d$-wave pairing parameters $d_{1,2}$ which, when nonzero, signal the presence of superconductivity and lead to a non-vanishing cluster $d$-wave order parameter
\begin{equation}
\psi = \sum_{\langle i,j\rangle_x} c_{i\uparrow}c_{j\downarrow} - \sum_{\langle i,j\rangle_y} c_{i\uparrow}c_{j\downarrow}
+ \mbox{c.c}.
\end{equation}

The order parameter is extracted from the lattice Green's function while the doping is measured on the cluster. The distance function is defined in Ref.~\onlinecite{Kancharla:2008}. We use cutoff $\om_c=2$ and fictitious inverse temperature $\beta=50$. 

While the extended interaction $V$ within the cluster can be treated exactly with an approach like CDMFT, the coupling to neighboring clusters requires the Hartree approximation.~\cite{AichhornEvertzEffectV:2004} More specifically, the model Hamiltonian
Eq.~\eqref{eq:hamil} is modified by replacing the inter-cluster interaction by
\begin{equation}\label{eq:hartree}
V \sum_{\langle i,j\rangle_\mathrm{c}}n_in_j  + 
V \sum_{\langle i,j\rangle_\mathrm{ic}}(\bar n_i n_j + n_i\bar n_j - \bar n_i \bar n_j)
\end{equation}
where $\langle i,j\rangle_\mathrm{c}$ denotes nearest-neighbor pairs within the plaquette and $\langle i,j\rangle_\mathrm{ic}$ nearest-neighbor pairs across plaquettes. The mean field $\bar n_i$ must be determined self-consistently; in practice, it is treated like the six bath parameters wihtin the CDMFT self-consistency loop and thus both the dynamical mean field (represented by the bath parameters) and the static Hartree mean field are converged simultaneously.

\begin{figure}
\centering{
\includegraphics[width=0.9\hsize]{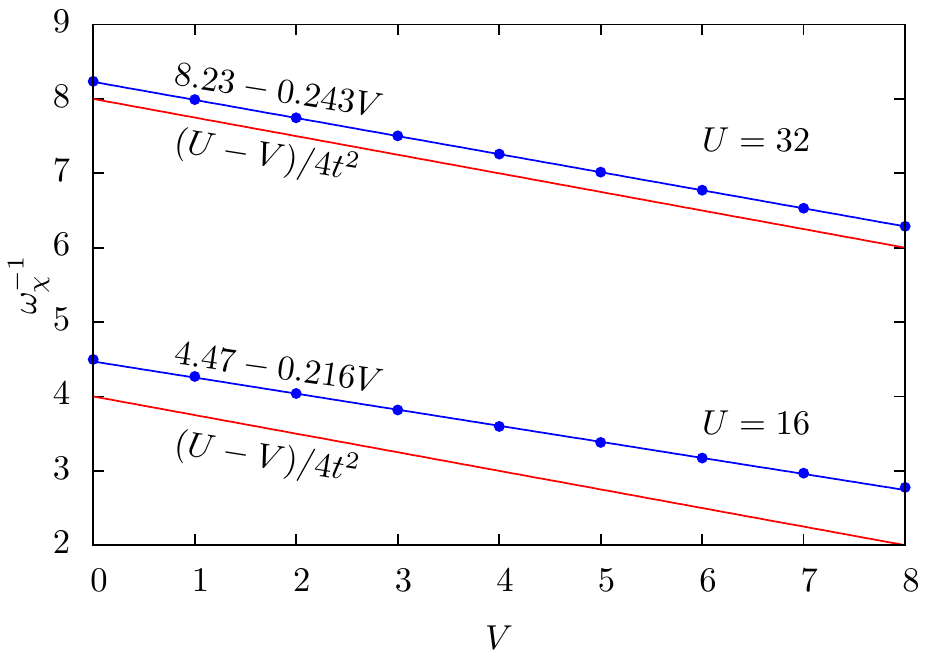}}
\caption{(Color online) Blue dots: Inverse frequency of the dominant peak in the antiferromagnetic susceptibility of the extended Hubbard model at half-filling, as a function of the nearest-neighbor interaction $V$, for $U=16$ and $U=32$ ($t=1$). These inverse frequencies are expected to be $J^{-1}=(U-V)/4t^2$ in the large-$U$ limit.}
\label{fig:chi_hf}
\end{figure}

To verify the accuracy of this approach, consider momentarily the half-filled, particle-hole symmetric extended Hubbard model with $t'=t''=0$.
In that simple case, the Hartree field $\bar n_i$ is fixed by particle-hole symmetry. We find that the dynamical spin susceptibility
at the antiferromagnetic wavevector $\Qv=(\pi,\pi)$ has a dominant peak at a low frequency $\om_\chi$. The position of this peak is shown on Fig.~\ref{fig:chi_hf} as a function of $V$ for two values of $U$. In the large $U$ limit, the half-filled extended Hubbard model should map to the Heisenberg model with a super-exchange parameter $J=4t^2/(U-V)$. The energy denominator is easily understood by comparing the energy, at $t=0$, between a configuration where all sites are exactly singly-occupied and another configuration in which one electron has vacated a site in order to doubly occupy a neighboring site. Four $V$ bonds are lost, but three are gained. Since we expect the frequency $\om_\chi$ to scale like $J$ in the large-$U$ limit, Fig.~\ref{fig:chi_hf} confirms that we obtain the correct scaling with $V$. Hence, our approach leads to the correct strong-coupling physics.

\begin{figure}
\centering{
\includegraphics[width=0.9\hsize]{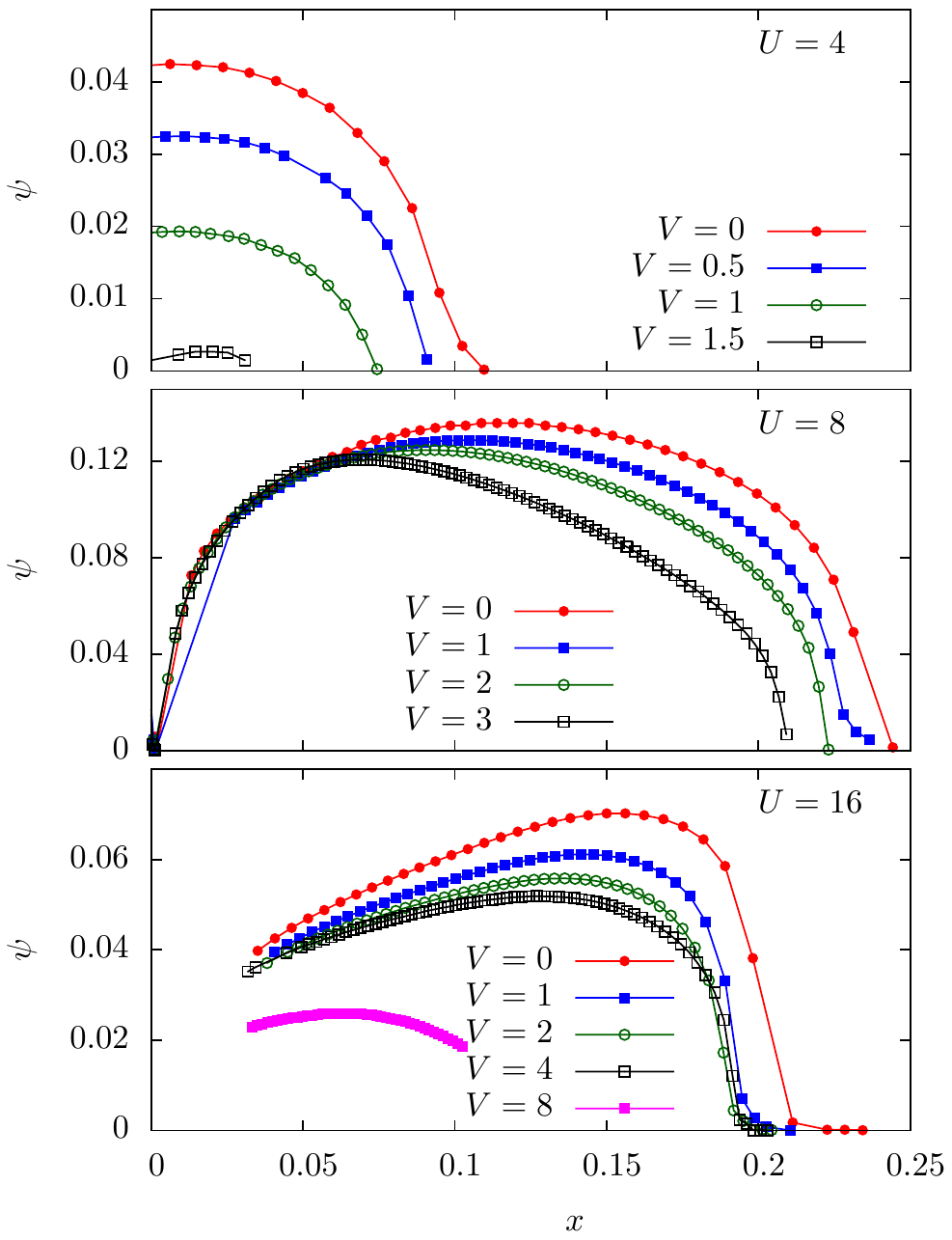}}
\caption{(Color online) $d$-wave order parameter $\psi$ obtained from the off-diagonal component of the lattice Green's function as a function of cluster doping for $U=4$, $8$ and $16$ and various values of $V$.}
\label{fig:OP}
\end{figure}

\section{Results}

 To concentrate on the effect of short-range spin fluctuations on superconductivity, we ignore the possibility of long-range antiferromagnetic order predicted by CDMFT at low doping,~\cite{Kancharla:2008} as well as the possibility of long-range charge density waves in the presence of $V$. At weak coupling charge-density waves may occur for $U\approx 4V$ in two-dimensions.~\cite{ZhangCallaway:1989,OnariExtended:2004,Davoudi:2007} We found a similar result at half-filling for strong coupling. 

The $d$-wave order parameter for YBCO hopping parameters is displayed on Fig.~\ref{fig:OP} for hole-doping at $U=4$, 8 and 16. In each panel, obtained for a given value of $U$, we display the results for different values of $V$. The Mott transition at half-filling, that separates weak from strong coupling, occurs around $U=6$.~\cite{Vekic:1993,park:2008,Balzer:2009} As observed before,~\cite{Kancharla:2008} in the weak coupling case ($U=4$) superconductivity is strongest at half-filling when competition with antiferromagnetism is prohibited (Fig.~\ref{fig:OP}a). At larger coupling, the Mott transition destroys superconductivity at half-filling (Figs.~\ref{fig:OP}b,c). 

The value of $V$ influences the order parameter in strikingly different ways at weak and at strong coupling. For $U=4$, superconductivity has essentially disappeared at $V/U=1.5/4=0.375$, in agreement with the upper limit $V/U=U/W$ found by weak coupling analysis.~\cite{Raghu:2012} This is consistent with the fact that at weak coupling, $V$ always decreases the strength of spin fluctuations.~\cite{ZhangCallaway:1989,Davoudi:2007} By contrast, at strong coupling, $V$ can increase the strength of spin fluctuations through $J=4t^2/(U-V)$, and for the same ratio $V/U=0.375$, the order parameter is still large for $U=8$ on Fig.~\ref{fig:OP}b . In fact, it is barely influenced by $V$ in the pseudogap (underdoped) regime close to half-filling ($x=0.05$). On the lower panel, for $U=16$, we have $V/J=16$ at $V=8$ and superconductivity still persists. The ratio $V/J$ is maximum at $V=U/2$. We estimate that $V\approx U/2$ is the upper bound for the the appearance of $d$-wave superconductivity (ignoring competing orders).  

\begin{figure}
\centering{
\includegraphics[width=0.8\hsize]{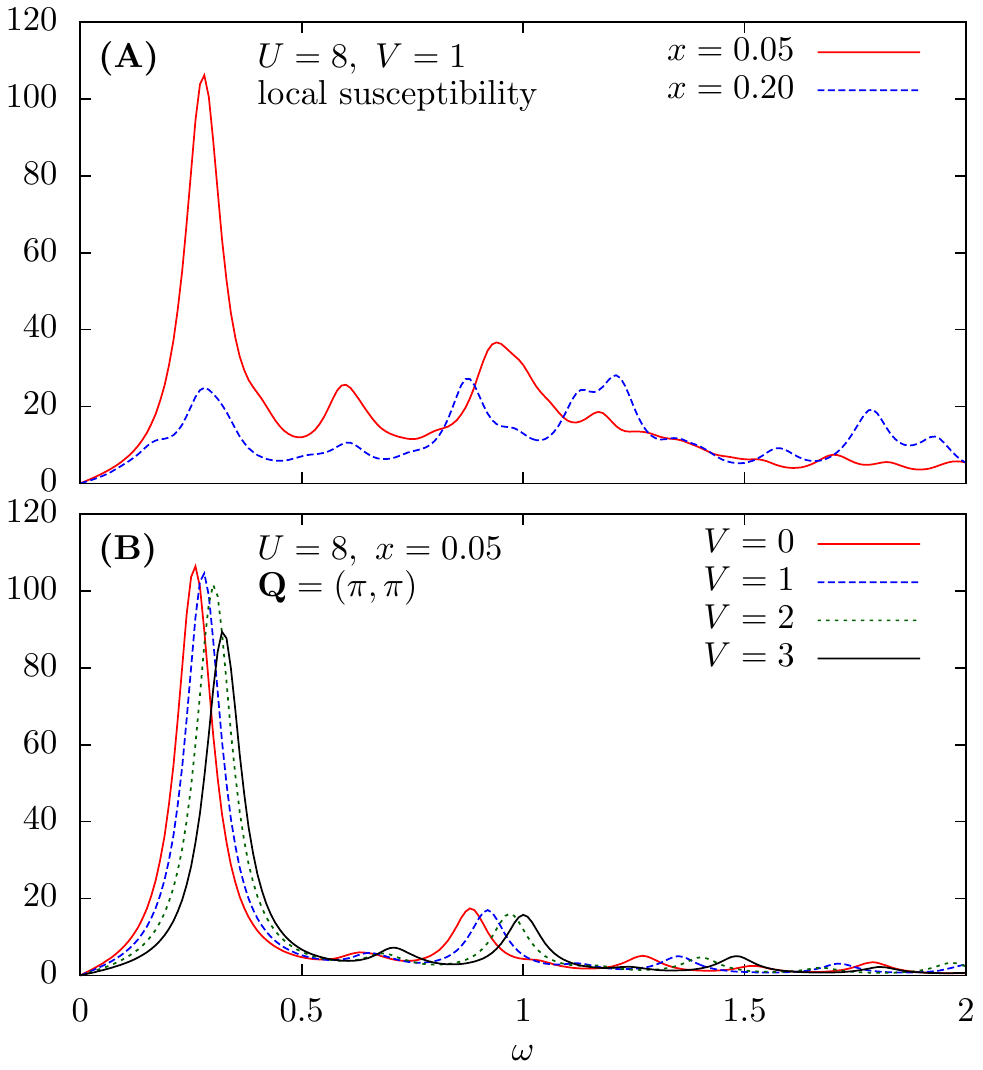}}
\caption{(Color online) (A) Imaginary part of the local spin susceptibility $\chi^{\prime\prime}(\om)$ for $U=8$, $V=1$ for an underdoped ($x=0.05$) and an overdoped ($x=0.20$) case. (B) In the underdoped regime ($x=0.05$) apart from a shift of the low-frequency peak to higher frequencies, it is mostly the amplitude, not the structure of the imaginary part of the plaquette antiferromagnetic spin susceptibility $\chi^{\prime\prime}(\Qv,\om)$ that is slightly affected by $V$.}
\label{fig:chi}
\end{figure}

The link between superconducting order parameter and spin fluctuations was demonstrated at $V=0$ by the correlation between the peaks in the imaginary part of the anomalous self-energy and the peaks in the spin fluctuation~\cite{Kyung:2009,Maier:2008}. This correlation persists here (not shown). To further investigate this link between spin fluctuations and superconductivity, let us thus first focus on the spin dynamics revealed by the the local spin spectral function $\chi^{\prime\prime}(\om)$ illustrated on Fig.~\ref{fig:chi}.
Panel (A) shows $\chi^{\prime\prime}(\om)$ at $U=8$, $V=1$ for the underdoped ($x=0.05$) and overdoped ($x=0.20$) regimes. The corresponding charge susceptibility is negligible on this scale. Spin fluctuations are much smaller in the overdoped regime and their weight is spread in wavevector contrary to the underdoped case. Panel (B) shows that in the underdoped regime ($x=0.05$) where there is a pseudogap, a moderate extended interaction $V$ at strong coupling mainly shifts the spectrum of $\Qv = (\pi,\pi)$ spin fluctuations to higher frequencies, as expected from the increase of $J=4t^2/(U-V)$ caused by $V$; it decreases somewhat the overall spectral weight without affecting the spectral shape.

\begin{figure}[!htb]
\centering{
\includegraphics[width=0.8\hsize]{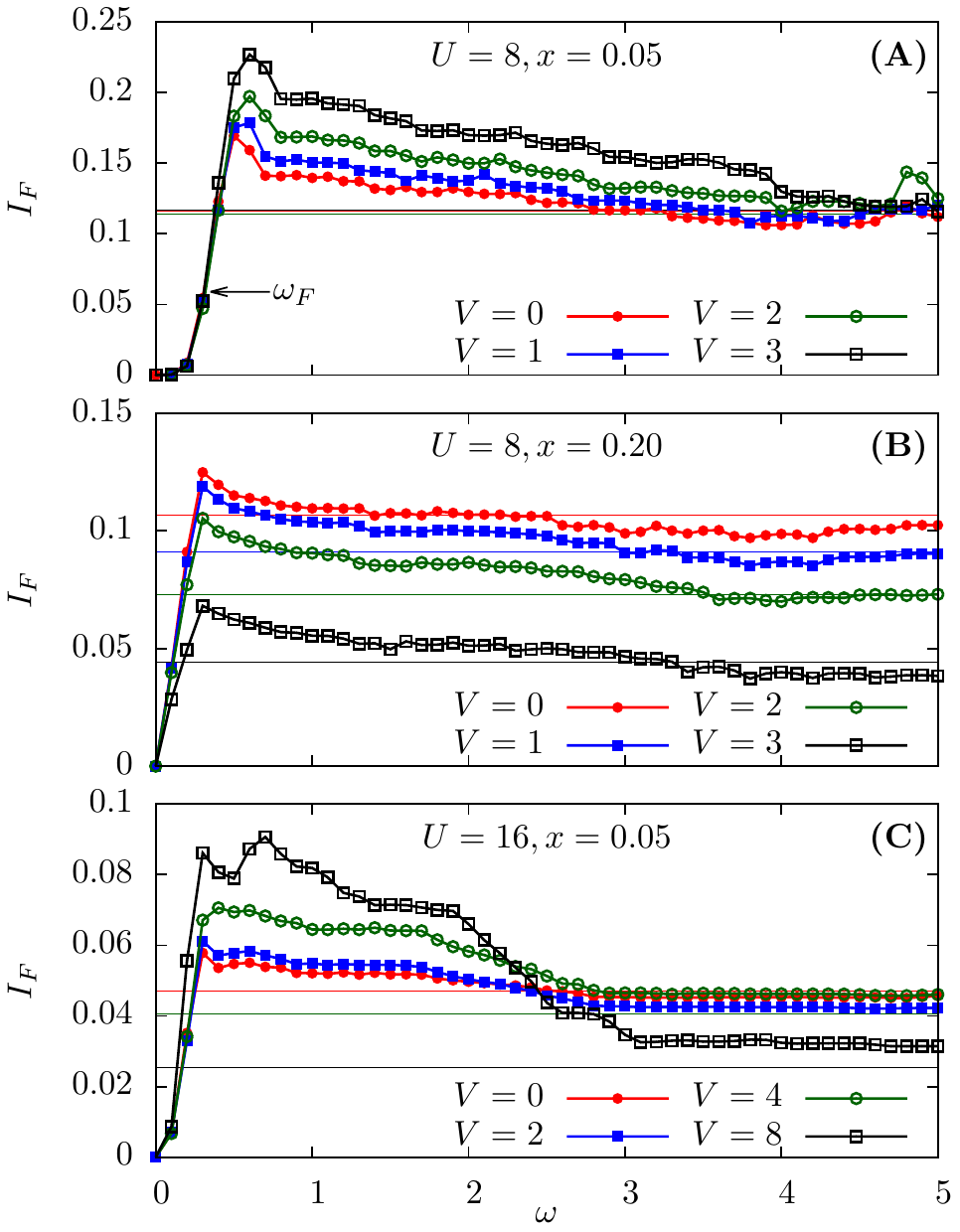}}
\caption{(Color online) Integral of the anomalous Green's function (or Gork'ov function) $I_F(\om)$ obtained after extrapolation to $\eta=0$ of $\om+i\eta$ for several values of $V$ at (A) $U=8$, $x=0.05$ (B) $U=8$, $x=0.2$ (C) $U=16$, $x=0.05$. The asymptotic value of the integral, $I_F(\infty)$, equal to the order parameter, is shown as horizontal lines. We call $I_F(\om)$ the cumulative order parameter. The characteristic frequency $\om_F$ is defined as the frequency at which $I_F(\om)$ is equal to half of its asymptotic value. The horizontal arrow in panel (A) indicates how $\om_F$ is obtained.}
\label{fig:IG}
\end{figure}

To show that low-frequency spin fluctuations are apparent in the pair dynamics, we compare the characteristic frequency $\om_\chi$ of spin fluctuations, given by the position of the dominant peak of $\chi^{\prime\prime}(\Qv,\om)$, with a characteristic frequency in the pair dynamics, as was done in Ref.~\onlinecite{Kyung:2009}.
The pair dynamics can be studied through the integral
\begin{equation}
I_{F}(\om) = -\int_0^\om\frac{\mathrm{d}\om'}{\pi}
\operatorname{Im}F_{ij}^{R}(\om')
\label{F(w)}
\end{equation}
where $F^{R}$ is the retarded Gork'ov function (or anomalous Green's function) defined in imaginary time by $F_{ij}\equiv-\langle Tc_{i\uparrow}(\tau)c_{j\downarrow}(0)\rangle$ with $i$
and $j$ nearest-neighbors. 
The infinite frequency limit of $I_{F}(\om) $ is equal to $\left\langle c_{i\uparrow}c_{j\downarrow}\right\rangle $ which in turn is proportional to the $T=0$ $d$-wave order parameter $\psi$.  
$I_{F}(\om)$ is useful to estimate the frequencies relevant for binding. We call $I_{F}(\om)$ the cumulative order parameter. 

	In BCS theory, $I_F(\om)$ is a monotonically increasing function of $\om$ that reaches its asymptotic value at the BCS cutoff frequency $\om_c$.~\cite{Kyung:2009} In the Eliashberg approach that includes retardation as well as the Coulomb pseudopotential,~\cite{Kyung:2009} the function overshoots its asymptotic value at frequencies near the main phonon frequencies before decaying to its final value because of pair breaking effects at higher frequencies. 
We define the characteristic frequency $\om_F$ as the point where $I_F(\om)$ reaches half of its asymptotic value.
If one imagines that $F_{ij}^{R}(\om)$ is made of a single peak, then that peak would be located at $\om_F$. Figure~\ref{fig:IG} illustrates the cumulative order parameter $I_F(\om)$ for various values of $V$. The top two panels are for two values of doping $x$ at $U=8$, and the bottom one for $U=16$ in the underdoped regime. The asymptotic value is indicated by a horizontal line. Following a sharp rise around $\om=\om_F$, the function has a maximum and then decreases towards its asymptotic value, as in Eliashberg theory.


\begin{figure}
\centering{
\includegraphics[width=0.9\hsize]{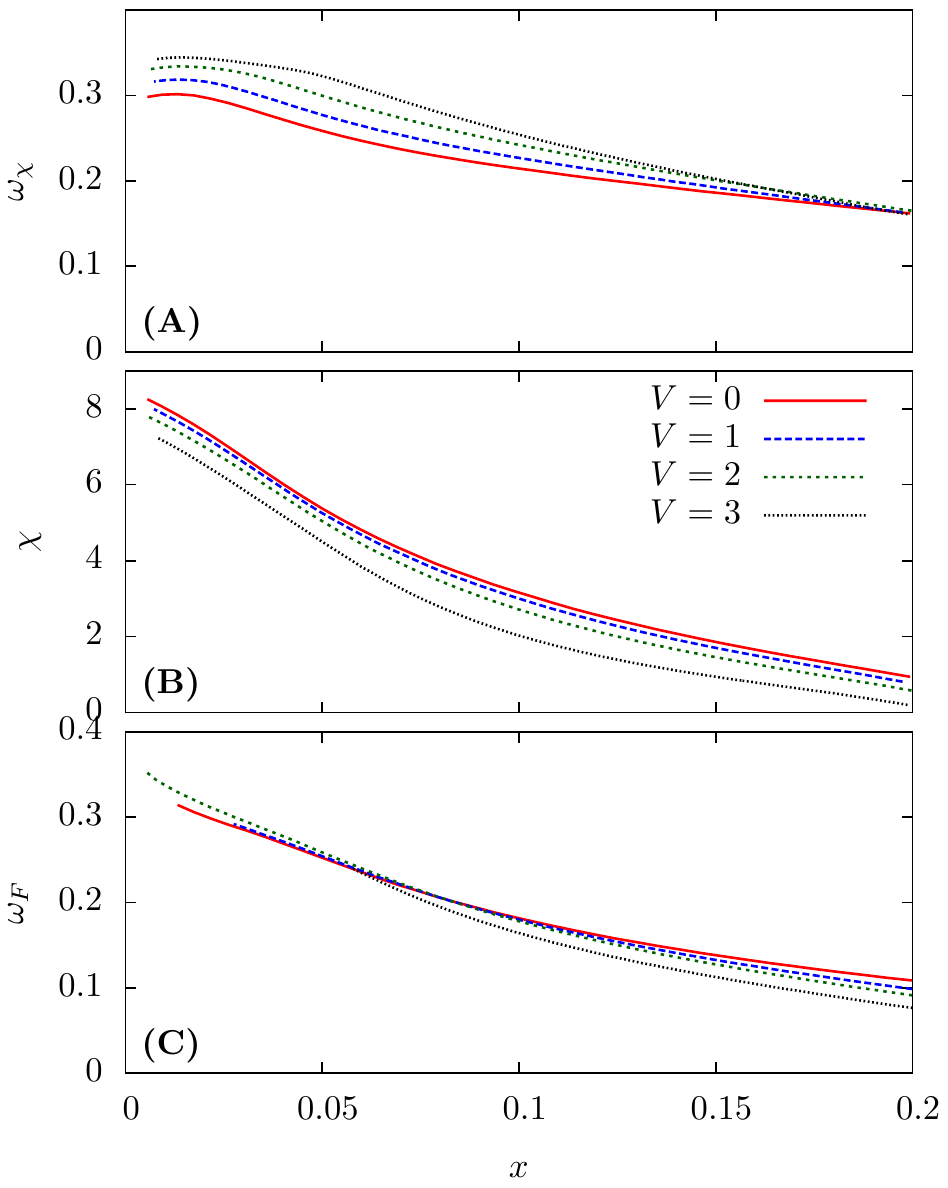}}
\caption{(Color online) Panel (A) position of the lowest peak in the AF susceptibility $\chi^{         \prime\prime}(\om)$ and the strength $\chi$ of that
peak on Panel (B), for $U=8$. Panel (C) shows the corresponding SC characteristic frequency $\om_F$.}
\label{fig:U8}	
\end{figure}

The link between the characteristic spin-fluctuation frequency $\om_\chi$ and the characteristic frequency of the pair $\om_F$ is summarized on Fig.~\ref{fig:U8}, which shows the evolution with doping of: (A) the position $\om_\chi$ of the main AF susceptibility peak, (B) its strength, and finally (C) the characteristic frequency $\om_F$, for four values of $V$ at $U=8$. Clearly, $\om_\chi$, $\chi$ and $\om_F$ all decrease with doping. This shows that spin fluctuations and pair dynamics are strongly linked.

\begin{figure}
\centering{
\includegraphics[width=0.9\hsize]{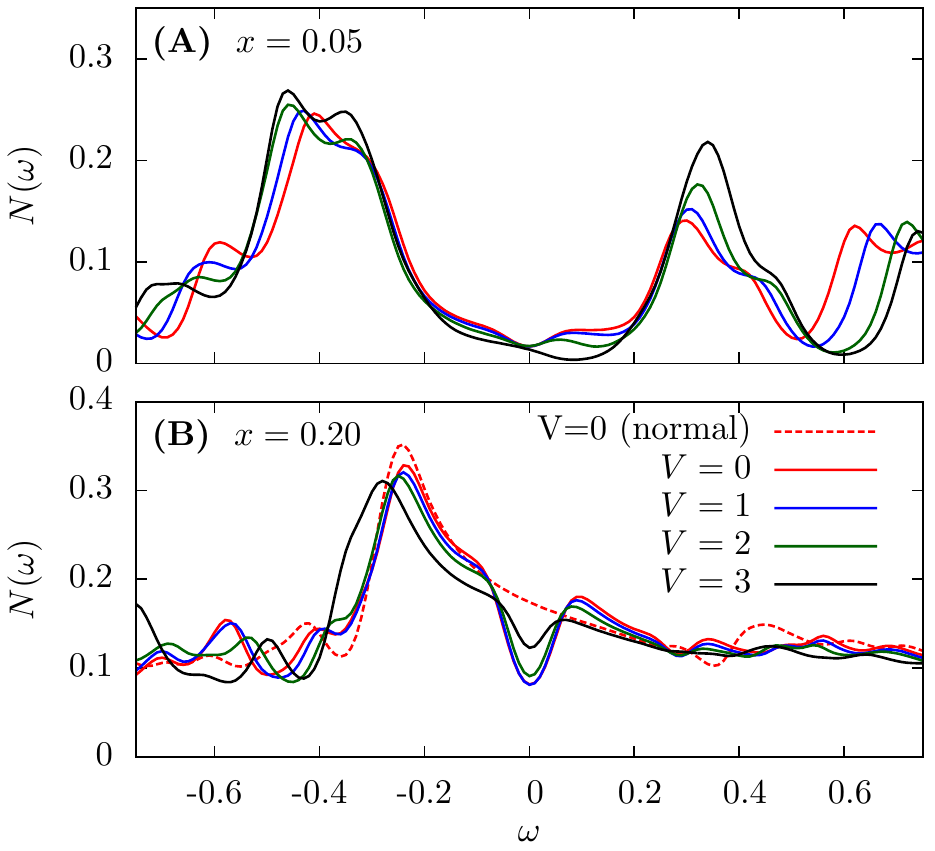}}
\caption{(Color online) Local density of states for four values $V=1,2,3,4$ and $U=8$ for (A) underdoped $x=0.05$, and (B) overdoped $x=0.2$ regimes. The dotted line in (B) is for the normal state $V=0$, all other lines in (A) and (B) are for the superconducting state.}   
\label{fig:DOS}	
\end{figure}
	
Even for $V=0$ however, the increase in $J$ as measured by $\om_{\chi}$ and $\om_F$ in the underdoped regime, $x<0.10$, does not correspond to an increase in the order parameter. As can be seen in Fig.~\ref{fig:OP}, for $x=0.05$ and $x=0.20$ the $d$-wave order parameter is essentially identical when $V=0$ and the increase in $\om_F$ near half-filling manifests itself only in the maximum value of the cumulative order parameter $I_F(\om)$ in Fig.~\ref{fig:IG}, not in the value of the order parameter. Clearly the order parameter does not increase in the underdoped regime despite the increase in the characteristic spin fluctuation frequency $\om_{\chi}$ because of another effect. That effect is the pseudogap. The pseudogap removes many of the states near the Fermi energy that would otherwise be paired. 

The presence of the pseudogap can be seen by contrasting the single-particle local density of states at two dopings. In the overdoped regime ($x=0.20$) the local density of states has no pseudogap in the normal state, as seen from the dashed red line on Fig.~\ref{fig:DOS}b. Only the superconducting gap is visible. By contrast, even at $V=0$ there is clearly a pseudogap in the underdoped regime, as can be seen from Fig.~\ref{fig:DOS}a. The effect of superconductivity manifests itself only through the small more symmetric gap near $\om=0$.~\cite{Aichhorn:2007} The pseudogap, as measured from the peak to peak distance, increases slightly with $V$ in the underdoped regime. 

The detrimental effect of the pseudogap on the superconducting order parameter was demonstrated in Fig.~7 of Ref.~\onlinecite{Kyung:2009} where the bubble contribution to the pairing susceptibility decreases as we approach half-filling. This means that in the absence of interactions between the particles forming the pair (represented by vertex corrections), self-energy effects disfavor superconductivity. This physics is also present at weak to intermediate coupling when the pseudogap is induced by long wavelength antiferromagnetic fluctuations~\cite{Kyung:2003} (and not by the incipient short-range Mott localisation that appears at strong coupling~\cite{Senechal:2004,Hankevych:2006}).

\section{Discussion}	

The effect of spin fluctuations and of near-neighbor repulsion $V$ on pairing is relatively straightforward in the overdoped regime ($x\gtrsim 0.10$). There, as seen on Fig.~\ref{fig:OP}b and Figs.~\ref{fig:U8}a,c, the order parameter increases with the characteristic spin frequency $\om_\chi$ and the characteristic pair frequency $\om_F$. Like for the $V=0$ case, superconductivity disappears at finite $V$ when the amplitude $\chi$ of the lowest frequency peak in the spin fluctuations (Fig.~\ref{fig:U8}b) vanishes,~\cite{Kyung:2009} as observed in experiment.~\cite{Wakimoto:2004,Wakimoto:2007} The main effect of $V$ is to decrease the spin fluctuations, as happens systematically at weak coupling.~\cite{ZhangCallaway:1989,Davoudi:2007} 	
	
 	As we approach half-filling, the decrease of the order parameter for all values of $V$ (Figs.~\ref{fig:OP}b,c), despite the increase in the amplitude $\chi$ of the lowest peak in the spin fluctuations (Fig.~\ref{fig:U8}), is mainly due to the increase of the pseudogap: it removes more and more states from the Fermi energy as the doping $x$ decreases (Fig.~\ref{fig:DOS}). Note that $V$ causes opposite changes in the amplitude $\chi$ and in the characteristic frequencies (Figs~\ref{fig:U8}), which might explain the near independence of the order parameter with respect to $V$ in Figs.~\ref{fig:OP}b,c. However, these changes are small on a relative scale and are probably not the main reason for the insensitivity of the order parameter to $V$. 
 	
 	The resilience of $d$-wave superconductivity to $V$ at strong coupling is best understood from the $U=16$ results for the cumulative order parameter $I_{F}(\om)$ shown in Fig.~\ref{fig:IG}c for $x=0.05$. The largest value of $I_{F}(\om)$ scales roughly like $J=4t^2/(U-V)$, in other words $V$ increases the binding at low frequency, where the retardation is large. However, $V$ also has the expected pair-breaking effect: larger $V$ causes larger pair-breaking so that the low frequency increase in $I_F(\om)$ is essentially compensated by the time $I_F(\om)$ reaches its asymptotic value (the order parameter).

	To understand more deeply the dual role of $V$, as both pair binding and pair breaking, it is helpful to return to the solution of the Eliashberg equations in the electron-phonon case. Figs.~7, and 13 of Ref.~\onlinecite{Scalapino:1966} show that the imaginary part of the gap function remains positive for all frequencies when the Coulomb pseudopotential vanishes. By contrast, one verifies from Fig.~11 of the same paper that the imaginary part of the gap function changes from positive to negative as frequency increases when the Coulomb pseudopotential is finite. A sign change in the imaginary part of the gap function should lead to a similar behavior in the imaginary part of the Gorkov function $F(\om)$. In this case its integral $I_F(\om)$, as defined in Eq.~(\ref{F(w)}), should reach a maximum before decreasing to its asymptotic large frequency value. By contrast, when the imaginary part of the gap function remains positive at all frequencies, i.e. in the absence of Coulomb pseudopotential, $I_F(\om)$ should increase monotonically to its asymptotic value, as in BCS theory. Since we observe that $I_F(\om)$ has a maximum at finite frequency, we are in the case where the Coulomb pseudopotential is important. Comparison of Figs.~11 and 13 of Ref.~\onlinecite{Scalapino:1966} also shows that the maximum value of the imaginary part of the gap function (and that of its frequency integral, as one can check) increases with the electron-phonon coupling constant. In our case, the increase in $J=4t^2/(U-V)$ caused by $V$ similarly leads to an increase in the maximum value of the cumulative order parameter $I_F(\om)$. In other words, we can surmise that the increase in $V$ leads to larger binding at low frequency, like an increase in the electron-phonon coupling constant, but it also leads to more pair-breaking at large frequency, coming from the Coulomb pseudopotential-like effect of $V$, and these two effects nearly cancel each other.        	

	In the electron-phonon case, one also notes that the imaginary part of the gap function vanishes at about two to three times the maximum phonon frequency. This is where the cumulative order parameter $I_F(\om)$ would reach its asymptotic value. In the case of spin fluctuations, $2J$ is a measure of the width of the spin-fluctuation spectrum. Comparing Fig.~\ref{fig:IG}a for $U=8$ and Fig.~\ref{fig:IG}c for $U=16$, one notices that the frequency for which  $I_F(\om)$ reaches its asymptotic value is smaller by a factor of roughly two for $U=16$ where the $V=0$ value of $J$ is twice as small as for $U=8$. 
 	
	 	
	The opposing effects discussed above conspire to leave the order parameter rather insensitive to $V$ in the presence of a pseudogap. The pseudogap induced by Mott physics~\cite{Sordi:2012} is detrimental to superconductivity, but in its presence superconductivity is effectively protected from near-neighbour repulsion $V$. Indeed, in a doped Mott insulator, short-range incipient localisation is strong enough to create a pseudogap, but while $V$ causes pair-breaking at high frequency, it also enhances spin fluctuations ($J=4t^2/(U-V)$) at low frequencies, thus compensating the pair-breaking effect.
Overall, retardation effects are crucial for the resilience of $d$-wave superconductivity to near-neighbor repulsion.	
	
\acknowledgments
	
We are indebted to D.J. Scalapino, A. Chubukov, J. Carbotte and G. Kotliar for useful discussions and to Giovanni Sordi for a careful reading of the manuscript. This work was supported by NSERC (Canada), CFI (Canada), CIFAR, and the Tier I Canada Research chair Program (A.-M.S.T.). Computational facilities were provided by Compute Canada and Calcul Qu\'ebec.


\end{document}